\documentclass[amssymb,twocolumn,aps]{revtex4}
\usepackage{graphics,color,graphicx,amsmath}
\usepackage{subcaption}
\usepackage{natbib}
\usepackage{verbatim}
\usepackage{graphicx}
\usepackage[above,below]{placeins}
\usepackage{float}
\usepackage{tabularx}
\usepackage{times}
\usepackage{blindtext}
\usepackage{url}
\usepackage{rotating}
\usepackage{mathtools}
\usepackage{threeparttable}

\begin{document}

\begin{titlepage}

  \title{Evaluating recognition and recall formats of social network surveys in physics education research}

  \author{Meagan Sundstrom$^{1}$, Justin Gambrell$^{2}$, Adrienne L. Traxler$^{3}$, and Eric Brewe$^{1}$}
    \affiliation{$^{1}$Department of Physics, Drexel University, Philadelphia, Pennsylvania 19104, USA\\
    $^{2}$Department of Computational Mathematics, Science and Engineering, Michigan State University, East Lansing, Michigan 48824, USA\\
    $^{3}$Department of Science Education, University of Copenhagen, Copenhagen, Denmark}

  % \keywords{}

  \begin{abstract}
    An increasing number of studies in physics education research use social network analysis to quantify interactions among students. These studies typically gather data through online surveys using one of two different survey formats: \textit{recognition}, where students select peers' names from a provided course roster, and \textit{recall}, where students type their peers' names from memory as an open response. These survey formats, however, may be subject to two possible systematic errors. First, students may report more peers' names on a recognition survey than a recall survey because the course roster facilitates their memory of their interactions, whereas they may only remember a subset of their interactions on the recall format. Second, recognition surveys may be subject to name order effects, where students are more likely to select peers' names that appear early on in the roster than those that appear later on (e.g., due to survey fatigue).
    Here we report the results of two methodological studies of these possible errors in the context of introductory physics courses: one directly comparing 65 student responses to recognition and recall versions of the same network survey prompt, and the other measuring name order effects on 54 recognition surveys from 27 different courses. We find that students may report more peer interactions on a recognition survey than a recall survey and that most recognition surveys are not subject to significant name order effects. These results help to inform survey design for future network studies in physics education research.
    \clearpage
  \end{abstract}
  %% Adding the `\clearpage` is the hack to make the title page.  In 2020, the proceedings is
  %% going to be double blind.  This change makes it so that we can programmatically remove the
  %% title page.  In the future, other blinding measures should be taken as well (for example,
  %% removing self-citations).  This is not needed in 2019.

  \maketitle
\end{titlepage}

\section{Introduction}

Social network analysis is an increasingly popular method among physics education researchers for analyzing relationships between students (e.g., with which peers students interact about the course material)~\cite{brewe2018guide,dou2019practitioner,traxler2022networksoverview}. This method allows us to visualize relationships among peers and quantify their impacts on student outcomes. Several studies, for example, have identified that students who interact with more peers about their physics course tend to perform better in the course and have a stronger sense of belonging~\cite{bruun2013talking,williams2019linking,dokuka2020academic}.

Network data is typically collected through online surveys that ask students to self-report their relationships with their peers~\cite{sundstrom2022interactions,commeford2021characterizing}. These surveys are administered in one of two formats: \textit{recognition} and \textit{recall}. Recognition surveys provide a course roster and ask students to select the names of peers with whom they have a particular relationship. Recall surveys ask students to type out the names of peers with whom they have a particular relationship from memory in an open response format. Both recognition and recall surveys have been used in prior physics education research (PER) studies, but possible systematic errors associated with each type of survey have yet to be examined in this context. 

Research in the field of sociology has identified two possible errors related to network survey format: (1) recall bias and (2) name order effects. With regard to recall bias, previous studies have found that people recall fewer names (in some cases, less than 20\%) than they recognize: survey respondents forget some of the people with whom they have a relationship during a recall prompt and only remember those people once shown their name in a recognition prompt~\cite{bahrick1975fifty,hammer1980social,hammer1984explorations,sudman1985experiments,sudman1988experiments,hlebec1993recall,brewer1993patterns,neyer1997free,brewer2000forgetting}. Respondents are also more likely to forget acquaintances than close friends and are more likely to forget people with whom they interacted for a briefer amount of time than people with whom they interacted for a more prolonged time period~\cite{sudman1988experiments}. Therefore, the number and strength of relationships that respondents report, and any networks formed from these reports, may differ dramatically between recall and recognition survey formats.

%Responses and, correspondingly, networks formed from roster and recall survey formats, therefore, may differ dramatically.

%On one hand, the recognition format reduces recall bias, in which students forget peers' names, by providing a list of names. At the same time, this format may inflate students' nominations due to the ease of selecting names.  On the other hand, leaving students to recall is likely more prone to issues of recall bias, where students forget peers' names and/or misspell their peers' names. The recall format may also, however, reduce the possibility of nomination inflation because it takes more cognitive effort to recall a peer's name. Given these possible effects, it is important to understand the impacts of these two survey methodologies in order to lay the groundwork for purposeful collection and analysis of social network data in our field.

Recognition surveys may also be subject to name order effects, where students are more likely to select peers' names that appear earlier on the provided roster (usually alphabetized) than those whose names appear later on~\cite{poulin2008methodological,marks2016relations,liu2024name}. Such effects may be due to survey fatigue (e.g., respondents stop reading any of the provided names by the time they get to the bottom of the roster) and/or respondents perceiving that they have satisfied the survey completion requirements after selecting a few names at the top of the roster~\cite{liu2024name}. %Name order effects have been studied among middle school students, with significant effects being identified in some cases~\cite{poulin2008methodological,marks2016relations}, but not others~\cite{liu2024name}. 
Name order effects may bias the results of analyses that quantify individual students' positions in social networks, for example by increasing the likelihood that students whose names appear earlier on the roster are centrally positioned in the network as compared to students whose names appear later on.

As network studies become more common in PER, it is critical that we examine these possible systematic errors in the context of undergraduate physics courses. In this paper, we present the results of two methodological studies: (1) a comparison of 65 introductory physics students' responses to both a recognition and recall survey prompt probing the peers with whom they interacted and (2) measurements of name order effects on 54 recognition surveys asking introductory physics students to select the peers with whom they interacted. We aim to address the following research questions:
\begin{enumerate}
    \item To what extent does the number of peers students report interacting with vary between a recognition survey and a recall survey? (Study 1)
    \item How do peer interaction networks measured with recognition and recall surveys compare to one another? (Study 1)
    \item To what extent is the number of times a student is selected by their peers on a recognition survey correlated with their position on the provided roster? (Study 2)
\end{enumerate}

\section{Methods}

\subsection{Study 1}

We collected data from an introductory, calculus-based mechanics course at a large, private, Ph.D.-granting
institution in the northeastern United States with an R1 designation (Very High Research Spending and Doctorate Production)
according to the 2025 Carnegie Classification of Research Activity. 201 students were enrolled in the course, most of whom were first-year engineering students. The course took place in a large lecture hall and the instructor used Peer Instruction~\cite{mazur1997peer}. 

As part of two different research projects, we separately administered a recognition and a recall survey probing students' peer interactions. Both surveys were given online via Qualtrics during the last week of class. Most students completed the two surveys within a few days of each other. About half of the students completed the recognition survey before the recall survey, and the other half completed the recall survey before the recognition survey.

The recognition survey asked the following: ``Please choose from the list of people that are enrolled in your physics class the names of any other student with whom you had a meaningful interaction in class during the past week, even if you were not the main person speaking.'' Below the prompt was an alphabetized course roster with associated check boxes, from which students could select as many names as they wanted. Students completed this survey during class time (i.e., surrounded by their peers) and the whole survey took about five minutes to complete. 79\% of enrolled students (\textit{n} = 158) completed this survey.

The recall survey asked the following: ``Please list any students in your physics class that you had a meaningful interaction with about the course material this week.'' Instructions to find the course roster on the learning management system were also provided (we do not know how many students actually accessed the roster). Below the prompt were open text boxes where students could type in as many peers' names as they wanted. Students completed this survey prompt outside of class time (i.e., not surrounded by their peers). The prompt was part of a longer survey that took about 20 minutes to complete. 36\% of enrolled students (\textit{n} = 73) completed this survey. 

65 students completed both surveys.  We only analyze responses from these 65 students in order to make direct comparisons between the two survey formats.  

To address our first research question, we examined the distributions of the number of peer names reported by each student on each survey format. We did not conduct statistical tests because the survey response rates were low and may represent a biased subset of the full class population~\cite{nissen2018participation}; rather, this descriptive analysis offers preliminary results.

To address the second research question, we created two undirected peer interaction networks (one per survey format) using only the responses from the 65 matched respondents. We included all 201 students enrolled in the class in the networks because students could select any of their peers' names. Nodes in the networks represented students and undirected edges represented reported interactions between students (regardless of which student reported interacting with the other). We used undirected edges because we aimed to identify whether and how the results of a social network analysis may vary based on the survey format used; we did not aim to disentangle the effects of reporting a peer versus being reported by a peer. Treating the edges as undirected also reduces possible impacts of missing data because we assume an edge exists between two students even if only one of the students reported the interaction on the survey. We calculated four descriptive measures for each network:
\begin{itemize}
    \itemsep 0cm
    \item \textit{Density}: the number of observed edges as a fraction of the maximum number of edges,
    \item \textit{Transitivity}: the proportion of two-paths (e.g., student A reports a connection with student B and student B reports a connection with student C) that close to form triangles (e.g., student A also reports a connection with student C), 
    \item \textit{Number of components}: the number of groups of nodes that are connected to each other but
    not to any other nodes in the network, and
    \item \textit{Giant component size}:  the number of nodes contained in the largest component of the network.
\end{itemize}
We did not conduct any statistical tests on these measures; rather, we aimed to make preliminary comparisons. We emphasize that the survey response rates were low, especially when we only consider the matched respondents. This analysis does not provide a complete picture of the social networks in this physics course.

\subsection{Study 2}

As part of a large national research project, the same recognition prompt from Study 1 was administered on both pre- and post-semester surveys in 27 introductory physics courses at 24 different institutions across the United States. %All courses were mechanics or another first course in an introductory physics or astronomy course sequence, such that students had not formed social relationships within the discipline beforehand. 
The courses were taught by instructors who self-reported that they used one of four active learning methods: Peer Instruction~\cite{mazur1997peer}, Tutorials~\cite{mcdermott2002tutorials}, Investigative Science Learning Environment (ISLE)~\cite{etkina2007investigative}, and Student-Centered Active Learning Environment with Upside-down Pedagogies (SCALE-UP)~\cite{beichner2007student}.  Fourteen of the institutions are public and ten are private. Thirteen institutions are Ph.D.-granting, seven are Master's-granting, three are Bachelor's-granting, and one is Associate's-granting. Eight institutions are R1 (Very High Research Spending and Doctorate Production), three are R2 (High Research Spending and Doctorate Production), and four are RCUs (Research Colleges and Universities) according to the 2025 Carnegie Classification of Research Activity. Three institutions are minority-serving (Hispanic-Serving Institutions or women's colleges).  %The response rates likely do not impact the results of our analysis, described next, because every respondent had the option to select any of their peers' names. 

All 54 surveys (pre- and post-semester surveys in each of the 27 courses) had a response rate above 50\% of enrolled students (response rates ranged from 55\% to 100\%, with an average of 79\%). To address the third research question, we calculated the Pearson's correlation coefficient (and the corresponding \textit{p}-value to indicate statistical significance of the correlation coefficient), \textit{r}, between students' position on the alphabetized course roster and the number of peers who reported interacting with them for each of the 54 surveys. We examined the correlation coefficients by class size and survey timing (whether pre- or post-semester) to determine if any name order effects are related to these variables. For example, it is plausible that larger classes may be subject to larger name order effects because they have longer rosters.

\section{Results}

\subsection{Study 1: Students may report more interactions on a recognition survey than a recall survey}

%For the 65 matched students, 178 and 100 selections of peers were made on the recognition and recall surveys, respectively. Most selections made on the recall survey (69 out of 100) were also made on the recognition survey. Many of the selections made on the recognition survey (109 out of 178), however, were not also made on the recall survey. 

\begin{figure}[t]
    \centering
    \includegraphics[width=3.5in,trim={0 0.4cm 0 0 }]{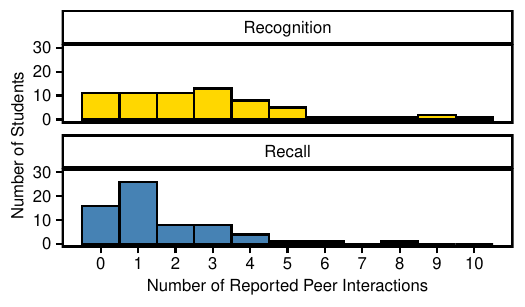}
    \caption{Distributions of the number of peers' names reported by each student on the recognition (yellow) and recall (blue) surveys (\textit{n} = 65).}
    \label{fig:RosterRecallBars}
\end{figure}

Students, on average, reported interacting with 2.7 (standard deviation, SD=2.3) and 1.6 (SD=1.6) peers on the recognition and recall surveys, respectively (Fig.~\ref{fig:RosterRecallBars}). The median number of peers reported by each student is two and one for the recognition and recall surveys, respectively. 33 students reported more peers on the recognition survey than the recall survey, 11 students reported more peers on the recall survey than the recognition survey, and 21 students reported the same number of peers on both surveys. Students, therefore, may report more peer interactions when given a recognition survey as compared to a recall survey.

%\begin{figure}[t]
 %   \centering
 %   \includegraphics[width=3.5in, trim={0.3cm 3cm 0 0.3cm }]{RosterRecallNetworks.pdf}
   % \caption{Interaction networks from the recognition (left, red) and recall (right, blue) surveys (\textit{N} = 65 respondents). Nodes represent students and edges represent reported interactions.}
   % \label{fig:RosterRecallNetworks}
%\end{figure}

\begin{figure}[t]
  \centering
  \subfloat[Recognition]
  {\includegraphics[width=1.65in,trim={3cm 2.5cm 2cm 2cm},clip]{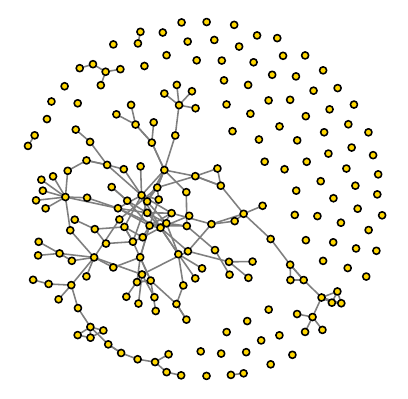}\label{fig:net1}}
  \hfill
  \subfloat[Recall]{\includegraphics[width=1.65in,trim={3cm 2.5cm 2cm 2cm},clip]{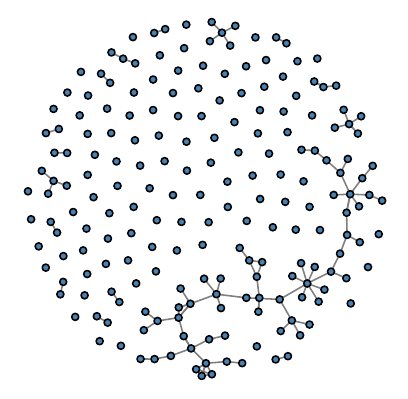}\label{fig:net2}}
  \caption{Interaction networks from each survey only using reported interactions from the 65 matched respondents. Nodes represent all enrolled students (\textit{n} = 201) and edges represent reported interactions. \label{networks}}
\end{figure}

%Our analysis of the peer interaction networks formed by student responses further illustrates how the two survey formats may lead to different study results. 
The peer interaction networks corresponding to the recognition and recall survey responses also have distinct structures (Figs.~\ref{fig:net1} and~\ref{fig:net2} and Table~\ref{networkstatsappendix}). The recognition network contains many more edges (i.e., higher density) and has a higher tendency for small-group clustering (i.e., higher transitivity) than the recall network. These features can be seen in the highly interconnected structure of the recognition network (e.g., with many students connected to one another in the giant component of the network), as compared to the more sparse recall network (e.g., with one long chain-like structure across the bottom right of the network, a few small disconnected components around the perimeter, and many isolated nodes). The recognition network also has fewer total components than the recall network and a giant component double the size of that of the recall network, signifying more connectivity among the nodes. These different network structures highlight that researchers' choice of survey format may have a profound effect on the results of a network study. Recognition surveys seem to elicit more selections of peers' names and, correspondingly, may result in more interconnected networks than recall surveys.

\subsection{Study 2: Most courses do not exhibit significant name order effects on a recognition survey}

\begin{figure*}[t]
    \centering
    \includegraphics[width=7in, ,trim={0 0.4cm 0 0}]{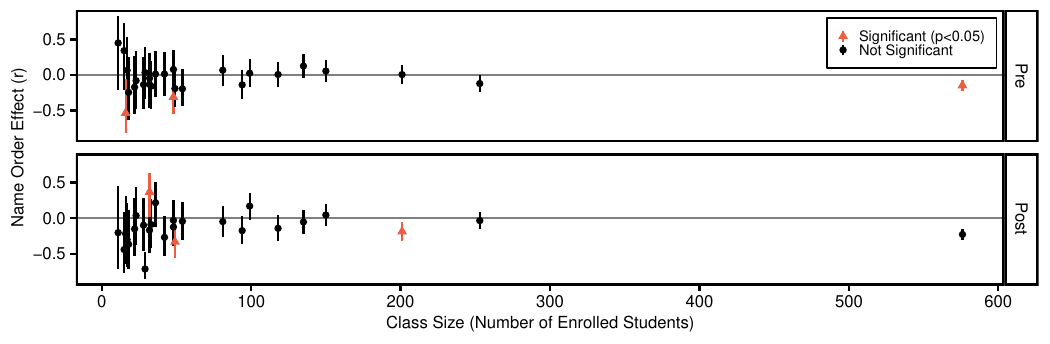}
    \caption{Correlation coefficients measuring the relationship between students' position on the alphabetized course roster and the number of peers who selected their name for 54 recognition surveys. Positive (negative) correlations indicate that students whose names appear later on in the roster are selected by more (fewer) peers. Error bars indicate 95\% confidence intervals.}
    \label{fig:NameOrder}
\end{figure*}

%Figure~\ref{fig:NameOrder} shows the name order effects (i.e., correlation coefficients) for 54 recognition surveys. 
For 48 of the 54 recognition surveys, there is no significant correlation between students' roster position and how many of their peers selected their name on the survey (black circles in Fig.~\ref{fig:NameOrder}). For five of the 54 surveys, there is a significant name order effect where students whose names appeared later on in the roster received fewer selections from their peers (red triangles with negative correlations in Fig.~\ref{fig:NameOrder}). For one survey, there is a significant name order effect where students whose names appeared later on in the roster received more selections from their peers (red triangle with positive correlation in Fig.~\ref{fig:NameOrder}). The surveys with significant name order effects are not systematically large or small courses (the red triangles span a range of class sizes) or pre- or post-semester surveys (three red triangles occur on pre-semester surveys and three red triangles occur on post-semester surveys). 

\begin{table}[t]
  \caption{Descriptive measures for the two observed networks (Figs.~\ref{fig:net1} and ~\ref{fig:net2}). The number of edges is lower than the total number of reported interactions (Fig.~\ref{fig:RosterRecallBars}) because undirected edges do not count reciprocal interactions separately.\label{networkstatsappendix}}
  \begin{ruledtabular}
    \begin{tabular}{lcc}
 & Recognition &  Recall \\ 
 \hline
 Nodes & 201 & 201  \\ 
 Edges & 160 & 89  \\ 
 Density & 0.009 & 0.005  \\ 
 Transitivity &  0.19 &  0.12 \\ 
Number of components &  5 &  17 \\ 
 Giant component size &  111 &  55 \\ 
    \end{tabular}
  \end{ruledtabular}
\end{table}

\section{Discussion and Conclusion}

We conducted two methodological studies related to network survey format in the context of undergraduate physics courses. In the first study, we found that students may report more peer interactions on a recognition survey as compared to a recall survey, even when these surveys probe the same type of relationship. These differences correspond to variability in the structure of social networks produced from student responses, with recognition surveys producing denser networks than recall surveys. Similar to prior work in other contexts~\cite{bahrick1975fifty,hammer1980social,hammer1984explorations,sudman1985experiments,sudman1988experiments,hlebec1993recall,brewer1993patterns,neyer1997free,brewer2000forgetting}, introductory physics students may report both strong (e.g., long and/or recurring interactions) and weak (e.g., short and/or infrequent interactions) relationships on a recognition survey, whereas students may only report strong relationships on a recall survey. It is important to note that neither survey format is ``correct;'' rather, researchers should identify the survey format that is best suited for their goals when designing a study. 

%Future research experiments should examine the extent to which nominations made on a recall survey but not a recognition survey are random versus systematic. For example, do students systematically ``forget'' their interactions with peers of a certain gender identity or racial/ethnic identity? This methodological question is important for making accurate measurements of gender and racial/ethnic effects within student social networks.

We acknowledge that there are several limitations to Study 1. First, there was a low percentage of enrolled students with matched responses, limiting generalizability. It is possible that our findings are mostly representative of high performers, who are more likely to fill out educational surveys than low performers~\cite{nissen2018participation}. Future research should conduct similar comparisons of recognition and recall surveys in other instructional physics contexts to validate our findings.

The two survey prompts were also not \textit{exactly} the same. The recognition survey asked students to report in-class interactions, while the recall prompt asked students to report both in- and out-of-class interactions. We might expect this difference to impact our results, however we saw fewer selections made on the recall survey despite this prompt eliciting more types of interactions than the recognition survey. %Our observations, therefore, are likely not attributable to the different types of interactions prompted by the two surveys.

Additionally, some students responded to the recall survey before the recognition survey, while others responded to the recognition survey before the recall survey. We do not believe the survey order impacted our results because about half of the students saw each survey first (as opposed to a majority seeing one survey before the other), yet the majority--54 of the 65 students--reported an equal number of peers or more peers on the recognition survey as compared to the recall survey. Future research experiments should test this possibility directly by controlling for the survey order and the time in between each survey, or by randomly assigning each student to respond to only one survey format.

We also could not completely isolate the effect of survey format on student responses in Study 1 because there were differences in total survey length and in whether the survey was completed during or outside of class time. The recognition survey was done in class, where students could look around and see their peers and remember to select them, possibly facilitating students in making survey selections. This survey was also shorter in total length, so students were likely not impacted by overall survey fatigue. The recall survey was done outside of class, where students had to do more cognitive work to remember their physics peers' names. This survey prompt also appeared later on in a longer survey, such that students may have experienced overall survey fatigue and, therefore, may have been less likely to report peers' names on this particular prompt.  Future research experiments should conduct more controlled comparisons of social network survey formats, for example using the same overall survey length and the same survey administration method.

In the second study, we observed that name order effects were not common on recognition surveys across 27 different introductory physics courses. Instead, students whose names appeared earlier versus later on the course roster had comparable probabilities of being selected, similar to one prior study~\cite{liu2024name}. This pattern likely indicates that students know their peers' names well enough to find them on an alphabetized course roster and that they do not read every name on the roster when making selections. Future research experiments, however, should compare student responses to recognition surveys with alphabetized versus randomly ordered rosters to further substantiate this claim.

There were also a few cases in Study 2 where name order effects were significant, as in previous work~\cite{poulin2008methodological,marks2016relations}. These effects were not systematic (i.e., by class size or survey timing) in our data and so may occur by random chance. We encourage researchers to check for such effects in their studies and, if necessary, make quantitative corrections. For example, researchers using students' number of peer interactions as a predictor variable in a regression model may add another predictor variable quantifying students' position on the alphabetized roster to account for name order effects.

Overall, the findings reported here indicate that the format of social network surveys has the potential to impact the results of PER studies. We recommend for researchers to carefully consider the types of relationships they wish to measure, make intentional decisions about their survey design, and account for systematic errors in quantitative analyses.

\acknowledgments{We thank the instructors and students who participated in our study. This work is supported by the National Science Foundation under Grant No. 2111128 and by the Cotswold Foundation Postdoctoral Fellowship at Drexel University.
}

\bibliography{bibfile} % don't include the .bib suffix

\end{document}